\documentclass[twoside,reqno]{qts-proc}
\usepackage{epsfig,cite}
\usepackage{amssymb,amsmath}
\usepackage{times}
\begin{document}
\sloppy \raggedbottom
\setcounter{page}{1}
%%%%%%%%%%%%%%%%%%%%%

\newpage
\setcounter{figure}{0}
\setcounter{equation}{0}
\setcounter{footnote}{0}
\setcounter{table}{0}
\setcounter{section}{0}


% use the thanks command within \title, \author or \address for footnotes:
% \title{Title} or  \title{Title\thanks{...}}

\title{Fractional space-like branes}

\runningheads{S. Kawai}{Fractional space-like branes}

\begin{start}

% \author{Name1}{aff.label1}, \coauthor{Name2}{aff.label2},  \coauthor{Name3}{aff.label3}
% \address{Address1}{aff.label1}\address{Address2}{aff.label2}\address{Address3}{aff.label3}

\author{Shinsuke Kawai}{1,2},
%\coauthor{Esko Keski-Vakkuri}{2,3},\\
%\coauthor{Robert G. Leigh}{4},
%\coauthor{Sean Nowling}{4}

\address{YITP, Kyoto University, Kyoto 606-8502, Japan}{1}
\address{Helsinki Institute of Physics, PL 64, Helsinki University, 00014 Finland}{2}
%\address{Department of Physical Sciences, PL 64, Helsinki University, 00014 Finland}{3}
%\address{Department of Physics, University of Illinois, % 1110W. Green Street, 
%Urbana, IL61801, USA}{4}

%you may repeat \coauthor as many times as you need
%names may have more than one aff.label, e.g.,
%\coauthor{Name2}{aff.label2,aff.label3},

\begin{Abstract}
We discuss construction and applications of instanton-like objects which we call fractional 
space-like branes. 
These objects are localised at a fixed point of a time-like (or more generally space-time) orbifold
which is a string theoretical toy model of a cosmological singularity. 
We formulate them in boundary state, adsorption, and fermionisation approaches.
% Presented at LMS Durham Symposium (July 2005, UK) and QTS-4 (August 2005, Varna, Bulgaria)
\end{Abstract}
\end{start}

%%%%%%%%%%%%%%%%%%%%%%%%%%%%%%%%%%%%%%%%%%%%%%%%%%%%%%%%%%%%%
% The main text of your paper                               %
%%%%%%%%%%%%%%%%%%%%%%%%%%%%%%%%%%%%%%%%%%%%%%%%%%%%%%%%%%%%%

\section{Fractional space-like branes in a time-like orbifold}

As a leading candidate of the ultimate theory of high energy physics, string/M-theory is expected to 
give physical explanations to highly conceptual issues such as the origin of the Universe. 
Indeed, in the past decade, much effort has been devoted to constructing consistent cosmological models
from a string theory perspective.
Various prototypical models, including Pre-BigBang\cite{Gasperini:2002bn}, Ekpyrotic\cite{Khoury:2001wf}, and Cyclic\cite{Steinhardt:2001st} scenarios, have emerged and attracted much attention. 
Although it would be fair to say that these models are rather immature, it is certainly worthwhile, and perhaps rewarding, to investigate string-based cosmological scenarios that could develop string theory into an experiment-backed science.  

These string-inspired cosmological scenarios typically possess big crunch / big bang singularities.
In these scenarios it is customary to assume that, through some resolution mechanism, the Universe undergoes a smooth transition from a big crunch to a big bang so that the history of the universe continues from the infinite past, evolving with bouncing event(s). 
This certainly poses various challenges. 
Besides the obvious problem of resolving the singularity, one often needs to implement extremely fine-tuned initial (asymptotic) conditions.
Since the age of such a universe is infinite, the second law of thermodynamics is likely to indicate an infinite amount of entropy today (i.e. a dead universe); in order to build a sensible scenario one needs to devise a mechanism to dump or dilute entropy.
If we compare these scenarios of string cosmology with the good old standard (big bang $+$ inflationary) cosmology, the unbounded past of the string-inspired models appears to be rather peculiar.
It is certainly not that the string theory 'forbids' cosmological models to have a finite age, but it is simply difficult to build a string theoretical background with non-trivial time dependence.
One obvious question that we may ask is whether it is possible to construct a finite-age cosmological model which is consistent with string theory.
In an attempt to do this, a space-time orbifold\cite{Balasubramanian:2002ry,Biswas:2003ku} has been proposed as a string toy model describing a space-time that is bounded in the past.
This is obtained simply by folding the Minkowski space-time with a ${\mathbb Z}_2$ identification, 
\begin{equation}
(X^0,X^i)\rightarrow (-X^0,-X^i),
\end{equation}
accompanied by the charge conjugation $C$. 
The number of spatial dimensions $d$ to be orbifolded ($i=1,\cdots,d$) is
$0\leq d \leq 25$ for bosonic and $0\leq d\leq 9$ for superstring.
The spacetime then includes fixed points of the orbifold action, and is not time-orientable
%%%
\footnote{Elliptic de Sitter space \cite{Parikh:2002py}, which can be embedded in the space-time orbifold, is also time non-orientable but is transitive with respect to the orbifold action.}.
It is of course natural to worry that this unhealthy-looking spacetime might not be compatible with string theory since strings are normally quite picky about the target space.
This issue has been studied in \cite{Balasubramanian:2002ry,Biswas:2003ku}.
Remarkably, it has been found that
(i) for certain $d$'s negative norm states are absent at tree level;
(ii) the usual closed string tachyon appears in bosonic theory but is absent in superstring theory;
(iii) there are no closed causal curves;
(iv) S-matrices are well defined so long as local properties are concerned.
These observations lead us to conclude that locally and away from the fixed points the spacetime orbifold retains good properties of the Minkowski space.

On the initial spacelike hypersurface (i.e. the fixed points of the time-like ${\mathbb Z}_2$ reflection) 
the one-loop vacuum expectation value of the energy-momentum tensor blows up so the spacetime is singular.
One may hope, in an analogy to the Euclidean orbifold case\cite{Diaconescu:1997br}, that such a singularity could be resolved by some mechanism involving (fractional) D-branes. 
In what follows we present our recent work\cite{Kawai:2005jx,KKLN} which could be considered as a study towards that direction.
We shall construct a new kind of spacelike branes, which we call fractional S-branes in an analogy to the Euclidean fractional D-branes, that sit at timelike orbifold fixed points.
These objects involve the twisted sector and are expected to capture some physics near an orbifold fixed point that models cosmological singularity. 

\section{Models of space-like branes and marginal boundary deformation}

Before discussing the orbifold theory, let us recall briefly the unorbifolded models of 
S-branes\cite{Gutperle:2002ai,Sen:2002nu,Larsen:2002wc}.
We focus on the spatially homogeneous case. 
The world sheet theory of S-brane models is realised by perturbing the string theory action by a boundary tachyon field, 
\begin{equation}
T(X^0)=\lambda_1\cosh \frac{X^0}{\sqrt{\alpha'}}+\lambda_2\sinh \frac{X^0}{\sqrt{\alpha'}},
\end{equation}
which is chosen such that upon Wick rotation the boundary tachyon vertex operators become exactly marginal, allowing the conformal invariance to persist for finite deformation. 
The tachyon profiles are classified into three equivalent classes depending on the values of 
$\lambda_1$ and $\lambda_2$. 
When $\lambda_1^2-\lambda_2^2>0$ (the elliptic equivalence class), the tachyon can be taken as
$T(X^0)=\lambda\cosh (X^0/\sqrt{\alpha'})$ by redefining the variables using the translational symmetry of $X^0$. 
This profile, called "full S-brane" in the literature, describes a tachyon which rolls up from one side of the potential hill, reaches a maximum at $X^0=0$ and then rolls down back to the original position at large $X^0$. Note that this profile is symmetric under the time reflection $X^0\leftrightarrow -X^0$.  
When $\lambda_1^2-\lambda_2^2=0$ (the parabolic equivalence class), using the symmetry we may rewrite the tachyon as
$T(X^0)=\lambda\exp(X^0/\sqrt{\alpha'})$.
This profile is called "half S-brane" and corresponds to a tachyon that starts from the top of the potential at $T(X^0=-\infty)=0$ with infinitesimal displacement and rolls down the hill (the side being determined by the sign of $\lambda$) at large $X^0$.
When $\lambda_1^2-\lambda_2^2<0$ (the hyperbolic equivalence class) the tachyon can be written as 
$T(X^0)=\lambda\sinh(X^0/\sqrt{\alpha'})$ by reparameterisation and the profile corresponds to a tachyon that starts from the bottom on one side of the potential, passes through the top at $X^0=0$ with finite velocity and rolls down the potential to the other side. 
The elliptic and hyperbolic cases are considered as creation and decay process of an S-brane, whereas in the parabolic case the S-brane only decays.

For constructing S-branes on the spacetime orbifold background we shall start from the full S-brane profile above since it has the time reflection symmetry $X^0\leftrightarrow -X^0$. 
We take this profile in the covering space and identify $X^0$ and $-X^0$ on the orbifold 
${\mathbb R}^{1,d}/{\mathbb Z}_2$. 
On the orbifold without boundary tachyon perturbation there are known to be two kinds of D-branes: 
{\em bulk} and {\em fractional}. 
The bulk branes exist away from the fixed points. 
These are D-branes symmetrised with respect to the ${\mathbb Z}_2$ action.
At the fixed points the D-branes consist of the twisted as well as the untwisted sector; these are called fractional D-branes.
In the next section we discuss dynamics of these branes with the boundary tachyon perturbation included.  

%%%%%
\begin{figure}[b]
\centerline{\epsfig{file=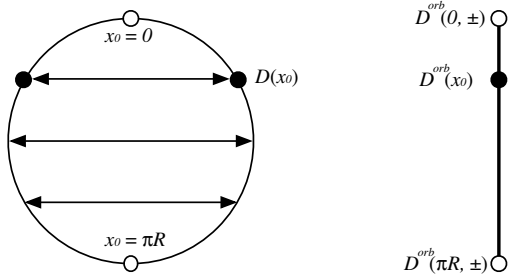,width=70mm}}
\caption{D-branes on $S_1$ (left) and on $S_1/{\mathbb Z}_2$ (right).
A D-brane on the bulk of $S_1/{\mathbb Z}_2$ (filled dot on the right) is the symmetrised superposition of the two corresponding D-branes on the covering space $S_1$. 
D-branes on the fixed points (the fractional branes) consist of twisted and untwisted sectors.}\label{figure:orbD}
\end{figure}
%%%%%

\section{World sheet descriptions of fractional S-branes}

The boundary dynamics of the $c=1$ orbifold theory without marginal deformation is discussed in
\cite{Oshikawa:1996dj,Cappelli:2002wq}. 
We work, in this section, in the Euclidean picture by Wick rotating the time as $X^0=iX$.
We first consider a circle $S^1$ of radius $R$ as the target space. 
One may place, at any position $x_0$ on the circle ($x_0\sim x_0+2\pi R$), a D-brane $D(x_0)$ that satisfies the Dirichlet boundary condition on $X$.
In the T-dual picture this D-brane is mapped to a Neumann brane, with a Wilson line taking its value $\theta_0$ on a circle of radius $\alpha'/R$.
We denote this Neumann brane as $N(\theta_0)$.
On ${\mathbb Z}_2$ orbifolding the target space becomes a line element of length $\pi R$ for the Dirichlet branes, so the value of $x_0$ is restricted to $0\leq x_0\leq \pi R$.
Except at the endpoints of the line element a D-brane of the orbifold theory is simply a symmetrised superposition of the two covering space D-branes, that is, in the boundary state language, 
\begin{equation}
\vert D(x_0)\rangle^{orb} = 2^{-1/2}\vert D(x_0)\rangle + 2^{-1/2}\vert D(-x_0)\rangle.
\end{equation}
Such a bulk D-brane cannot be moved over to a fixed point as it does not correspond to an irreducible representation anymore at the fixed points.
Instead, one needs to introduce fractional D-branes, 
\begin{equation}
\vert D(\xi,\pm)\rangle^{orb} = 2^{-1/2}\vert D(\xi)\rangle \pm 2^{-1/4}\vert D(\xi)\rangle^{tw},
\end{equation}
where $\xi$ means either of the end points $\xi=0, \pi R$, and the superscript $tw$ means the twisted sector.
The coefficients are determined by consistency. 
Thus there are four fractional D-branes, $D(0,+)$, $D(0,-)$, $D(\pi R,+)$ and $D(\pi R,-)$ (see Fig.1).  
Likewise in the T-dual image the value of the Wilson line in the orbifold theory is restricted to $0\leq\theta_0\leq\pi\alpha'/R$.
A Neumann brane with $\theta_0\neq 0, \pi\alpha'/R$ is a symmetrised superposition of two Neumann branes on the covering space,
\begin{equation}
\vert N(\theta_0)\rangle^{orb} = 2^{-1/2}\vert N(\theta_0)\rangle + 2^{-1/2}\vert N(-\theta_0)\rangle,
\end{equation}
and the fractional Neumann-branes with the end-point Wilson line, $\zeta=0,\pi\alpha'/R$, are
\begin{equation}
\vert N(\zeta,\pm)\rangle^{orb} = 2^{-1/2}\vert N(\zeta)\rangle \pm 2^{-1/4}\vert N(\zeta)\rangle^{tw}.
\end{equation}
Thus there are four fractional Neumann branes, $N(0,+)$, $N(0,-)$, $N(\pi\alpha'/R,+)$, $N(\pi\alpha'/R,-)$, so we have 8 fractional branes in total.
The question we are addressing now is what will happen to, say, $D(0,+)$, once we turn on the boundary tachyon field. 
In other words, we are studying the moduli space of the $c=1$ orbifold boundary theory for the particular deformation direction corresponding to our tachyon profile. 
Following Sen's interpretation\cite{Sen:2002nu}, this will allow us to analyse the dynamics of the fractional S-branes.
In the following subsections we will present three approaches for analysing such a deformation.

\subsection{Closed string formulation -- boundary states}

The boundary deformation of the $c=1$ orbifold CFT was first studied by Recknagel and Schomerus in \cite{Recknagel:1998ih}, using the boundary state formalism.
The key idea was to use the equivalence of two bosonic CFTs, one compactified on $S^1$ at $R=(1/2)\sqrt{\alpha'}$ and the other on $S^1/{\mathbb Z}_2$ at $R=\sqrt{\alpha'}$\cite{Ginsparg:1987eb}.
There is an underlying $SU(2)$ symmetry in the CFT of a free boson.
This symmetry is explicit when the boson is compactified on a circle of self-dual radius $R=\sqrt{\alpha'}$, which is just the WZNW model at level 1. 
The group is generated by three generators, $J_\pm=e^{\pm 2iX/\sqrt{\alpha'}}$ and $J_3=i\partial X$.
A part of the symmetry remains also in the two models of CFT which we are considering, and it can be shown that $J_1^{orb}=\cos (2X/\sqrt{\alpha'})$ of the orbifold at $R=\sqrt{\alpha'}$ corresponds to $J_3^{circ}=i\partial X$ of the circle theory at $R=(1/2)\sqrt{\alpha'}$.
Since the $J_3^{circ}$ shifts the position of a D-brane along the circle, we find that on the orbifold side the $J_1^{orb}$, which upon Wick rotation is nothing but the elliptic profile of the tachyon perturbation, gives rise to a deformation of boundary whose moduli space is $S^1$ in that direction.
This picture can be made more explicit by studying the partition functions. 
We find, as the parameter $\lambda$ varies, the 8 fractional boundary states change into one another,
\begin{eqnarray}
&&D(0,+) \rightarrow N(0,+) \rightarrow D(\pi\sqrt{\alpha'},+) \rightarrow N(\pi\sqrt{\alpha'},-)\rightarrow
D(0,-)\nonumber\\
&& \rightarrow N(0,-) \rightarrow D(\pi\sqrt{\alpha'},-) \rightarrow N(\pi\sqrt{\alpha'},+)\rightarrow
D(0,+) \rightarrow\cdots
\label{eqn:defseq}
\end{eqnarray}
The 8 fractional branes are evenly spaced 8 points on the $S^1$.
We may then interpret this sequence as describing the dynamics of fractional S-branes on a time-like orbifold at the self-dual radius. 
In the limit of decompactification one needs to project out unnecessary excitations along the line of \cite{Gaberdiel:2001zq}. 
We find that states corresponding to those at $\pi\sqrt{\alpha'}$ in the sequence of the deformation (\ref{eqn:defseq}) decouple in the decompactification limit so that the deformation reduces to 
\begin{eqnarray}
D(0,\pm) \rightarrow N(0,\pm).
\end{eqnarray}

\subsection{Open string formulation -- adsorption method}

An alternative approach is to use the adsorption method of \cite{Callan:1993mw,Callan:1994ub}
and modify it to include the orbifolding action.
This is an open string formulation so the effect of the boundary is taken into account by computing a trace over appropriate Chan-Paton factors.
Following \cite{Callan:1994ub} we shall perturb the Hamiltonian density of the free boson, 
\begin{equation}
{\cal H}_0 = -:(\partial X)^2: = :J_3^2: = \frac 13 :\vec{J}^2: = :J_1^2:
\end{equation}
by the perturbation term
\begin{equation}
\delta{\cal H} = \lambda J_1,
\end{equation}
which is localised on the boundary.
Completing the square, the effect of the perturbation reduces to a momentum shift. 
Including the orbifolding action
\begin{equation}
g: X\mapsto -X
\end{equation}
in the trace, one may check that the open string partition function agrees, after modular transformation, with the one obtained in the boundary state formalism above.
 
\subsection{Open string formulation -- fermionisation method}

The above two approaches are, strictly speaking, valid only at the self-dual radius and taking the decompactification limit $R\rightarrow\infty$ is somewhat subtle. 
One may instead resort to the fermionisation method of \cite{Polchinski:1994my} and modify it for our orbifold analysis.
This involves various technicalities which are elucidated thoroughly in \cite{KKLN}.
A merit of using this approach is that the model reduces to free fermions with masses localised on the boundary, so the system can be extended to any radius of the orbifold once the fermionisation has been accomplished.

\section{Nucleation and decay of branes at the origin of time}

Physical interpretation for the process described by the fractional S-branes which we have introduced above involves Wick rotating back to the Lorentzian manifold and computing number density and energy density. 
As in the standard case\cite{Lambert:2003zr,Gaiotto:2003rm} such information may be extracted from calculation of an overlap 
\begin{equation}
\langle V\vert B\rangle
\end{equation}
between an on-shell closed string state $V$ and the full boundary state of the fractional S-brane, 
\begin{equation}
\vert B\rangle = \vert B\rangle_{X^0}\vert B\rangle_{sp}\vert B\rangle_{gh},
\end{equation}
where the subscripts $sp$ and $gh$ stand for the spatial and ghost parts.
The computation for the untwisted sector is actually similar to the standard case\cite{Lambert:2003zr}, although some technical issues specific to the orbifold construction need to be carefully worked out.
In particular, in our orbifold case it is natural to take the Hartle-Hawking integration contour\footnote{In the fractional S-brane case the integration is in fact performed for a double Hartle-Hawking contour. The Hilbert space also needs to be doubled.} since there is no real past direction in time.
In the fundamental domain of the spacetime orbifold we may interpret this as a nucleation of an unstable brane at $X^0=0$ and its subsequent decay into closed strings.  
In the twisted sector we can compute similar overlaps, although it is not entirely straightforward to interpret the results. 
For example, if one computes the overlap of the fractional S-brane boundary state and the twisted sector vacuum, one obtains a result involving a delta function in the direction of imaginary time, that might have something to do with a resolution mechanism.
Although we do not have a clear conceptual interpretation of the twisted sector at the moment, this is obviously an interesting problem that deserves further investigation. 

% To start a subsection:
% \subsection{subsection heading type here}

\section{Concluding remarks}

We have described our construction\cite{Kawai:2005jx,KKLN} of fractional space-like branes with boundary tachyon field included.
The boundary conformal field theory naturally describes a nucleation and decay process of such an unstable brane.
It is important to understand the dynamics of unstable branes and their decay processes in string theory, 
and the rolling tachyon picture based on exact boundary conformal field theory solutions is an extremely powerful approach.
Our solution, the fractional S-branes, is a new type of exact solution which describes a brane on a spacetime with a spacelike singularity. 
We conclude by listing possible directions of future research which we think is of potential interest. 
(i) the marginal deformation of fractional S-branes can naturally be interpreted as brane decay from the origin of time, which may be seen as a toy model of an initial condition for the beginning of a spacetime. 
This is reminiscent of the no-boundary proposal of quantum cosmology\cite{Vilenkin:1982de,Hartle:1983ai}.  
Although our model is admittedly very simple and clearly more sophistication is needed in order to compare with realistic scenarios of the Universe, it could be considered as a first step towards a string version of quantum cosmology. 
In particular, as the fractional S-branes are supposed to decay into closed string gas, it might be considered as an initial condition for the string gas cosmology\cite{Brandenberger:1988aj}.
(ii) closely related to the above, one could hope that fractional S-branes can play some r\^ole in resolving a cosmological singularity.
In the case of a Euclidean orbifold the relation between fractional and wrapped branes is well known, and it is natural to expect a similar mechanism for the Lorentzian case.
It should be noted that, while in the closed string picture the background is singular, in the dual open string picture the initial condition of the spacetime is well defined in the form of boundary conditions of open strings. 
(iii) a correspondence between de Sitter space and conformal field theory, which is a Euclidean version of the more or less established AdS/CFT correspondence, is being advocated in the hope of providing a tool to investigate non-perturbative aspects of various systems including cosmology.
In this context fractional S-branes in the (elliptic) de Sitter space may serve as a key tool to investigate the holographic dual.

%%%%%%%%% FIGURES
%to do figures, use the following format:
%
%\begin{figure}[b]
%\centerline{\epsfig{file=figure.eps,width=70mm}}
%\caption{Caption type here}\label{choose label here}
%\end{figure}

%%%%%%%%%%%%%%%%%%%%%%%%%%%%%%%%%%%%%%%%%%%%%%%%%%%%%%%%%%%%%
% Doing Acknowledgement                                     %
%%%%%%%%%%%%%%%%%%%%%%%%%%%%%%%%%%%%%%%%%%%%%%%%%%%%%%%%%%%%%

\section*{Acknowledgments}

I thank Esko Keski-Vakkuri, Robert G. Leigh and Sean Nowling for enjoyable collaboration.
Support from JSPS (Research Fellowships for Young Scientists) is acknowledged.

%%%%%%%%%%%%%%%%%%%%%%%%%%%%%%%%%%%%%%%%%%%%%%%%%%%%%%%%%%%%%
% Doing references:                                         %
%%%%%%%%%%%%%%%%%%%%%%%%%%%%%%%%%%%%%%%%%%%%%%%%%%%%%%%%%%%%%

\end{document}